\documentclass[aps,prb,reprint,groupedaddress,superscriptaddress,nofootinbib]{revtex4-1}
\usepackage{mathrsfs}
\DeclareMathAlphabet{\mathpzc}{OT1}{pzc}{m}{it}
\usepackage{color}
\usepackage{graphicx}
\usepackage{dcolumn}
\usepackage{bm}
\usepackage{array}
\usepackage{multirow}
\usepackage{epstopdf}

\usepackage{amsmath}
\usepackage{tabulary}
\usepackage{rotating}
\usepackage{graphicx} 
\usepackage{lipsum} 
\usepackage{tabularx}

\usepackage{subfigure}
\usepackage{parskip}

\usepackage{tikz}
\usepackage{amsmath}
\usepackage{footmisc}

\usepackage{float}

\usetikzlibrary{shadings,patterns}
\usetikzlibrary{positioning, shapes.geometric}

\begin{document}
	\preprint{APS/123-QED}

\title{Machine-learning charge density for electronic materials and devices simulation} 
\author{Qingyun Zhang}
\thanks{These authors contributed equally to this work.}
\affiliation{State Key Laboratory of Quantum materials,School of Physical Science and Technology, ShanghaiTech University, Shanghai, 201210, China}
\author{Yang Hu} 
\thanks{These authors contributed equally to this work.}
\affiliation{State Key Laboratory of Quantum materials,School of Physical Science and Technology, ShanghaiTech University, Shanghai, 201210, China}
\author{Aixia Zhang}
\affiliation{State Key Laboratory of Quantum materials,School of Physical Science and Technology, ShanghaiTech University, Shanghai, 201210, China}
\author{Zehua Sun}
\affiliation{State Key Laboratory of Quantum materials,School of Physical Science and Technology, ShanghaiTech University, Shanghai, 201210, China}
\author{Jiawei Yan}
\affiliation{2020 X-LAB, Shanghai Institute of Microsystem and Information Technology, Chinese Academy of Sciences, Shanghai 200050, China}
\author{Xiaohao Zhou}
\affiliation{State Key Laboratory of Infrared Physics, Shanghai Institute of Technical Physics, Chinese Academy of Sciences, Shanghai 200083, China}
\author{Wei Lu}\email{luwei@shanghaitech.edu.cn}
\affiliation{State Key Laboratory of Quantum materials,School of Physical Science and Technology, ShanghaiTech University, Shanghai, 201210, China}
\affiliation{State Key Laboratory of Infrared Physics, Shanghai Institute of Technical Physics, Chinese Academy of Sciences, Shanghai 200083, China}
\author{Youqi Ke}\email{keyq@shanghaitech.edu.cn}
\affiliation{State Key Laboratory of Quantum materials,School of Physical Science and Technology, ShanghaiTech University, Shanghai, 201210, China}


\date{\today}

\begin{abstract}

We present a machine-learning (ML) charge density model, DeltaChargeNet, and its integration with the full-potential tight-binding muffin-tin orbital (FP-TB-MTO) method for electronic structure and quantum transport simulations of materials and devices. In this approach, the ML model directly predicts the electron density from atomic structures with low linear-scaling cost, while the FP-TB-MTO method provides an accurate tight-binding representation of the system Hamiltonian.
To achieve high-fidelity ML density for complex device structures with interfaces and disorder, we train an equivariant graph convolutional network to predict $\Delta \rho$—the difference between DFT and atomic superposition densities—under a charge neutrality constraint necessary for inhomogeneous material systems. A secondary ML residual charge model is introduced to compensate for errors in the primary $\Delta \rho$ model. The predicted ML density is then used within the full-relativistic FP-TB-MTO framework combined with non-equilibrium Green's function method. We validate our method on bulk Hg$_{1-x}$Cd$_{x}$Te disordered alloys (including Hg/Cd vacancy defects) and Au/Hg$_{1-x}$Cd$_{x}$Te/Au disordered device structures for varying $x$. The maximum error in the band energy for the bulk is below 10 meV for bulk Hg$_{1-x}$Cd$_{x}$Te.  For the tested device structures, the charge neutrality constraint suppresses long-wavelength charge bias that would otherwise distort the electrostatic potential by 0.68–1.88 eV near the device center, and the residual compensation model corrects the remaining 0.16–0.46 eV potential mismatch in the central region, essential for producing the correct potential barrier profile. With both, the ML-based layer-resolved potential energy profile, including the interface potential steps and the bulk-like plateau, agrees closely with fully self-consistent DFT, with a maximum deviation below 0.1 eV. As a further demonstration, we compute quantum transport through an infrared detector structure Au/[CdTe/Hg$_{1-x}$Cd$_{x}$Te/CdTe]/Au comprising 11220 atoms. By combining first-principles accuracy with inference-time linear scaling, DeltaChargeNet enables the simulation of large-scale device structures by bypassing the full DFT self-consistent field loop.

\end{abstract}

\pacs{}

\maketitle

\section{Introduction}
As materials design and device simulations progressively extend into the nanoscale and mesoscale regimes, the demand for accurate electronic structure descriptions in systems of increasing size and complexity has grown substantially. The relentless miniaturization of electronic devices to the nanoscale has ushered in an era where quantum mechanical effects play a vital role in device behavior. Understanding and predicting these effects, particularly quantum transport, is critical for advancing next-generation technologies such as nano-transistors, quantum computing device structures, and ultra-sensitive photodetectors and sensors. Kohn–Sham density functional theory (KS-DFT) has become the workhorse for describing electronic structures and modeling quantum transport from first principles. However, KS-DFT calculations require many cycles of self-consistent field (SCF) solution of the Kohn–Sham equations, posing fundamental bottlenecks for systems containing thousands to millions of atoms—such as those involving defects, interfaces, heterostructures, amorphous materials, and realistic device structures—due to their prohibitive computational cost, challenging even with linear-scaling methods. This limitation severely restricts the rational design of large-scale electronic devices and materials, thus necessitates new paradigms beyond conventional SCF approaches..

Recent years have witnessed great efforts to integrate machine learning (ML) with first-principles calculations to achieve order-of-magnitude reduction in computational cost while maintaining high accuracy. These approaches span a broad spectrum, including machine learning interatomic potentials (MLIPs),\cite{Behler2007, Wang2018, Batzner2022, Batatia2022, Zhang2024, zhanglinfeng} Hamiltonian matrix elements,\cite{Schutt2019, Zhang2022, Li2022, Zhong2023, Gu2024} and direct prediction of system properties,\cite{Xie2018, Chen2019, Choudhary2021} thereby bypassing self-consistent field (SCF) calculations. However, each of these strategies carries inherent limitations. MLIPs typically do not provide direct access to electronic-level information such as band structures or quantum transport properties, while ML Hamiltonian approaches must pay special attention to the long-range interactions and boundary conditions (in solving Poisson equation) in complex structures such as interfaces and device structures.\cite{Kosmala2023, Zhong2026} In these inhomogeneous systems, interfacial charge redistribution gives rise to dipoles whose electrostatic influence is inherently long-ranged and can substantially affect band alignment, local barrier profiles, and device electrostatics.\cite{Monti2012,Zou2025} 

Among the various ML strategies, methods that target the electron density—the fundamental quantity in density functional theory via the Hohenberg–Kohn theorem\cite{Hohenberg1964}—are particularly attractive, as they offer a direct route to bypass costly SCF cycles while preserving the full information content of DFT.\cite{Grisafi2018, Fabrizio2019, Dick2020, Jorgensen2020, Jorgensen2022, Lewis2021, Lee2022, Cheng2023, Koker2024, Kim2024, Fu2024, Pathrudkar2024, Lee2024} The electron density satisfies the nearsightedness principle,\cite{Kohn1996} which makes it well suited for machine learning. Moreover, a density-based strategy provides direct access to the electrostatic potential and offers a physically consistent route to capturing long-range electrostatic responses induced by interfacial charge transfer. A successful ML model for density could, in general, feed into established first-principles frameworks to compute forces, total energies, Hamiltonians, band structures, and quantum transport with high fidelity.

However, achieving this vision requires overcoming significant challenges. First, the ML model must predict a high-dimensional, spatially complex field—the electron density—with high accuracy across diverse chemical environments. Second, it must respect fundamental physical constraints, including rotational equivariance (to ensure consistent predictions under symmetry operations) and charge conservation (which is particularly critical for inhomogeneous systems such as interfaces and heterostructures, where local charge neutrality is essential). Third, the predicted density must be seamlessly integrable with an electronic structure method to efficiently and reliably yield the Hamiltonian and final physical properties of interest. Presently, due to limited accuracy, insufficient transferability, or the lack of proper physical constraints, existing ML density methods remain largely restricted to relatively uniform bulk systems, while device structures containing multiple interfaces and disorders remain challenging. In particular, current ML density models cannot yet deliver the precision required for quantum transport simulations of device structures, where errors in the potential energy on the order of hundreds of meV or even exceeding 1.0 eV (as shown in Sec.\ref{resultsec}) can qualitatively alter transport characteristics.

In this work, we present an accurate ML density model called DeltaChargeNet, and its integration with the full-potential tight-binding muffin-tin orbital (FP-TB-MTO) method for electronic structure and quantum transport simulations of materials and devices, that achieves the first-principles accuracy. In this approach, a primary equivariant graph convolutional network (GCN) model is trained to predict the high-fidelity electron density difference, $\Delta \rho = \rho_{DFT} - \rho_{ref}$, directly from atomic structures, by satisfying the intrinsic physical symmetries and charge conservation. A secondary residual compensation model is introduced to correct the errors of the primary $\Delta\rho$, which is important for correctly describing interfacial dipoles and device electrostatics. The ML density is then fed into the first-principles framework of full-potential screened spherical wave based tight-binding muffin-tin orbital (FP-TB-MTO) method to construct system's Hamiltonian for the subsequent electronic structure and quantum transport simulation of device structures. FP-TB-MTO provides an accurate and minimal tight-binding representation of the system's Hamiltonian, making it a computationally efficient partner for the ML density model. We demonstrate the high efficiency and DFT-level accuracy of ML density model for electronic structure and quantum transport properties for Hg$_{1-x}$Cd$_{x}$Te disordered bulk alloys and  Au/Hg$_{1-x}$Cd$_{x}$Te/Au disordered device structures. 

  \begin{figure*}
   \centering
   \includegraphics[width=0.8\linewidth]{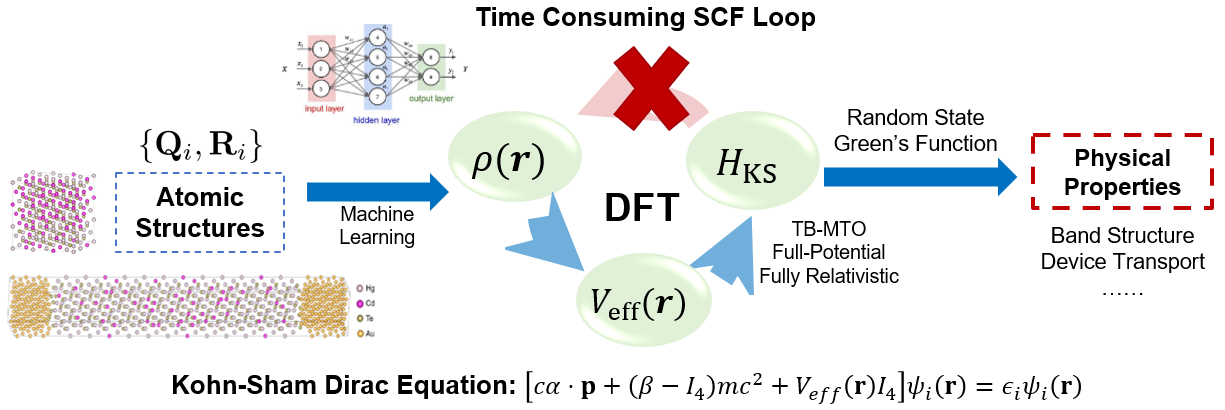}
   \caption{Workflow of ML density model accerlated simulation of materials and devices.}\label{workflow}
 \end{figure*}

\section{Methodology} 
 The computational workflow illustrated in Fig. \ref{workflow} outlines the ML-density accelerated scheme for first-principles simulations of large-scale materials and nanoscale devices. As the starting point, a ML density model, trained on the DFT density dataset of small systems,predicts an accurate charge density $\rho(\vec{r})$ from the atomic structure of a material or device. 
This ML density serves as an efficient alternative to the conventional self-consistent Kohn–Sham (KS) loop, thereby bypassing the iterative update between the density $\rho(\vec{r})$, the effective potential $V_{eff}(\vec{r})$, and the KS Hamiltonian. With  $\rho(\vec{r})$, the effective potential 
$V_{eff}(\vec{r})$ is computed by solving the Poisson equation over the entire simulation domain and calculating the exchange-correlation potential part. Based on the potential $V_{eff}(\vec{r})$, the full potential tight-binding muffin-tin orbital method is adopted to construct the KS Hamiltonian and overlap matrices of the system, providing an efficient representation of the electronic structure. At last, physical properties of large materials and devices, including band structures for bulk materials and quantum transport for electronic devices, can be calculated by combining with other techniques. 

\subsection{Machine-Learning Electron Density}

\subsubsection{Primary Model}
To avoid learning the full density $\rho_{DFT}(\vec{r})$, we decompose the density as \cite{Dick2020, Lee2024}
\begin{equation}
    \rho_{DFT}(\vec{r})=\rho_{ref}(\vec{r})+\Delta\rho(\vec{r})
\end{equation}
where the reference density $\rho_{ref}(r)$ is constructed as a superposition of atomic densities (SAD), namely
\begin{equation}
   \rho_{ref}(\vec{r})=\sum_{R} \rho^{Q}_{atom}(r_{R})
\end{equation}
where $\rho_{atom}^{Q}(r_{R})$ ($r_R = |\vec{r}-\vec{R}|$) denotes the known radial atomic density of element Q for the atom at site $\vec{R}$. The range of residual $\Delta\rho$  represents the charge redistribution induced by chemical bond formation, and is smaller than that of the total density, serving as the machine learning target. Here $\Delta\rho(\vec{r})$ satisfies the charge conservation condition
\begin{equation}
  \int \Delta \rho(\vec{r})d\vec{r} = 0.
\end{equation}

Predicting $\Delta \rho$ confines model errors to the physically informative part of the density, while the dominant background contribution $\rho_{ref}$ is provided deterministically. Moreover, $\Delta\rho$ has a much smaller dynamic range than $\rho$ reducing the dominance in MSE-type losses and allowing the optimization to focus on subtle but physically critical redistributions (due to bonding, defects, and interfaces). As a result, the ML reconstructed total density $\rho_{pred}=\rho_{ref}+\Delta\rho_{pred}$ is typically more accurate and numerically robust than learning $\rho$ directly.

Motivated by the requirements of device-scale $\Delta \rho$ learning, we developed a dedicated implementation and training framework for mapping periodic atomic graphs to atom-centered Gaussian orbital coefficients under SO(3) equivariance with the GCN,\cite{Cheng2023}  following the general idea of coefficient-based density representation.\cite{Grisafi2018, Fabrizio2019, Lewis2021, Lee2022,Cheng2023, Fu2024, Kim2024} The present formulation and implementation are designed specifically to ensure correctness, scalability, and stability for large heterogeneous systems. In the density reconstruction stage, $\Delta \rho(\vec{r})$ is written in a periodic multicenter form by including lattice translations $\mathbf{T}$ in the relative displacement,
\begin{equation}\label{eq:Gaussian_orbital_density}
  \Delta\rho_{prim} = \sum_{\mathbf{B}, \mathbf{T},L,\alpha} c_{\mathbf{B},\alpha,L}\cdot r_{\mathbf{B}\mathbf{T}}^{l} \exp(-\alpha r_{\mathbf{B}\mathbf{T}}^{2})Y_{L}(\hat{\vec{r}}_{\mathbf{B}\mathbf{T}})
\end{equation}
\color{black}
where $L=lm$ and the coefficient $c_{\mathbf{B},\alpha,L}$ encodes the information of atomic graphs,  $\vec{\mathbf{B}}$ denotes the site in the unit cell and $\vec{r}_{\mathbf{BT}} \equiv \vec{r}-(\vec{\mathbf{B}}+\vec{\mathbf{T}})$.

To achieve an accurate ML model for large device heterostructures, several methodological aspects are particularly important. 
First, we establish a consistent treatment of periodic boundary condition (PBC). Accurate prediction of $\Delta\rho$ in periodic heterostructures requires PBC to be treated consistently in both graph construction and real-space density reconstruction. In the atomic graph, neighbors and edge vectors are constructed using a minimum-image convention tailored for cubic cells. This provides well-defined PBC-aware relative vectors for equivariant message passing without explicitly replicating periodic images. 
Second, the atom-centered Gaussian exponents are arranged in a geometrically spaced sequence, so as to cover the relevant interatomic length scales while remaining compatible with the real-space grid resolution used in this work. This choice balances coverage over multiple length scales with numerical efficiency in large heterogeneous systems.
Third, in addition to pointwise supervision of $\Delta\rho$, we include a global electron-number conservation term to stabilize training across structures of different sizes. Rather than estimating the total charge by summing $\Delta \rho$ on the real-space grid, we compute it directly from the monopole ($l=0$) coefficients of the Gaussian expansion. The predicted charge deviation is
\begin{equation}
   \Delta Q_{pred} = \sum_{\mathbf{B},\alpha} w_{\alpha,l=0} \cdot c_{\mathbf{B},\alpha,l=0}
\end{equation}
where $w_{\alpha,l=0}$ are the normalization factor of the $l=0$ radial Gaussian basis in Eq.\ref{eq:Gaussian_orbital_density}. The loss function is then defined as
\begin{equation}
   \mathcal{L} = \mathcal{L}_{MSE} + \lambda (\Delta Q_{pred}/\Omega)^2
\end{equation}
where $\Omega$ is the cell volume, and $\lambda$ is selected to yield a stable trade-off between charge conservation and pointwise error. We find that imposing charge conservation constraints on the small systems within the training set improves local charge neutrality upon generalization to large systems, which is necessary for the correct potential-energy profile in device structures.
Finally, to make the real-space reconstruction  for large systems, the atom-centered Gaussian contributions are accumulated in fixed-size chunks over the expansion centers. This decouples the GPU memory footprint from the total number of atoms while leaving the Gaussian expansion itself unchanged. 

 \begin{figure*}[htbp]
     \centering
     \includegraphics[width=0.9\linewidth]{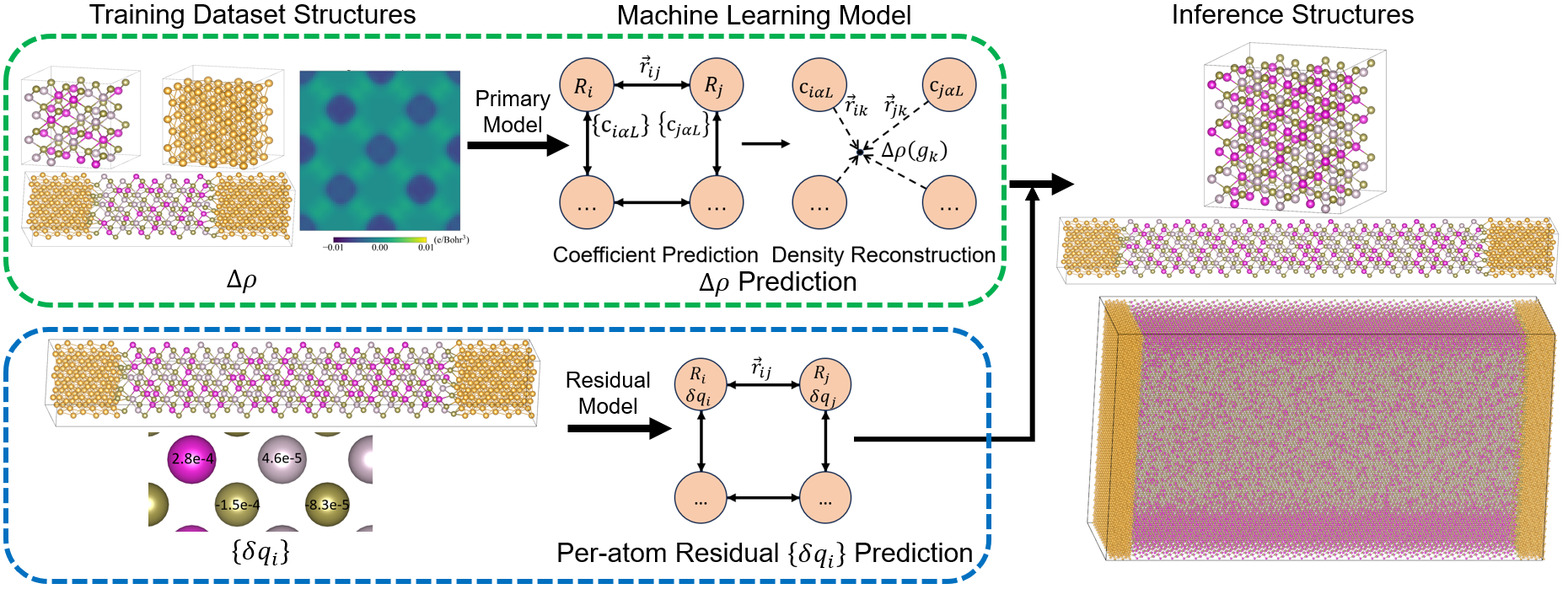}
    \caption{Workflow of ML-based charge density prediction.}
    \label{pred electron density fig1}
 \end{figure*}

\subsubsection{Residual compensation model}

Although the primary model already incorporates a charge-conservation constraint, this alone is not always sufficient to eliminate the residual long-wavelength charge bias in device-scale predictions. Even when the remaining density error appears small on the real-space grid, a weak long-wavelength charge bias distributed across atomic layers can still produce a noticeable error in the reconstructed potential, which is particularly important for downstream device calculations, as the potential is much more sensitive to such accumulated long-wavelength charge errors than to short-wavelength local density fluctuations.

To further correct this residual bias, we introduce a secondary residual-compensation ML model. Instead of directly learning the residual density on the grid, we project the density residual onto atom-centered scalar targets, so that the second-stage model only needs to predict the equivalent residual charge associated with each atom.

Concretely, after obtaining the primary model prediction $\Delta\rho_{prim}(\vec{r})$, we define the residual density on the uniform grid as
\begin{equation}
   \delta\rho(\vec{r})=\Delta\rho_{DFT}(\vec{r})-\Delta\rho_{prim}(\vec{r}),
\end{equation}
and convert it into an atom-wise equivalent residual charge
\begin{equation}
   \delta q_R=\int \frac{w_{R}(r_R)}{\sum_{R'} w_{R'}(r_{R'})}\delta\rho(\vec{r})\mathrm{d}\vec{r},
\end{equation}
where $w_{R}(r_R)=\exp \left(-r_R^2/2s^2\right)$ is a given distance-based partition weight function. We then train a dedicated lightweight correction GCN network to predict $\delta q_R$ from the atomic structure. These predicted charges are redistributed back onto the uniform grid through atom-centered compensation density,
\begin{equation}
   \begin{aligned}
      &\delta\rho_{res}(\vec{r})=\sum_R \delta q_R \phi_R(\vec{r}_R),\\
      &\phi_R(\vec{r}_R)\propto r_R \exp \left(-\frac{r_R^2}{2\sigma^2}\right),
   \end{aligned}
\end{equation}
and then the corrected difference density is finally written as
\begin{equation}
   \Delta\rho_{pred}(\vec{r})=\Delta\rho_{prim}(\vec{r})+\delta\rho_{res}(\vec{r}).
\end{equation}
Here, the residual-compensation network follows the same atom-centered message-passing philosophy as the primary model, but predicts atom-wise scalar charges rather than orbital coefficients.

From a practical standpoint, eliminating this residual correction using only the primary model would likely require more extensive long-channel training data and more message-passing layers. By focusing only on the small errors left after the primary prediction, the residual compensation model provides a more targeted and potentially more economical route to improving potential consistency in long-channel device structures.

\subsubsection{Model Inference}

At inference time, we exploit the locality of the learned operator to reduce the cost of evaluating the density on large device structures. The uniform grid is partitioned into spatial batches; for each batch, we extract a local atomic subgraph (via k-hop neighborhoods) that contains only atoms contributing to that region, and apply the model on this truncated subgraph to obtain the batch-wise $\Delta\rho$ prediction. Together with block-wise accumulation, the memory and computation are governed by the chosen local graph and batch size rather than the total system size.

To avoid repeated graph building, we precompute a global neighbor list once for the full structure and construct each mini-batch subgraph by filtering the cached list. The predicted total density is then reconstructed by adding back superposition of atomic density $\rho_{ref}$ of the corresponding structure,
\begin{equation}\label{full_model}
   \rho_{pred}(\vec{r}) = \rho_{ref}(\vec{r}) + \Delta\rho_{pred}(\vec{r}).
\end{equation}

Then the machine-learning charge density can be used as input to a first-principles DFT framework to calculate the physical properties of electronic materials and devices, bypassing the self-consistent DFT calculations.

\subsection{Fully-relativistic full-potential TB-MTO method}

\subsubsection{ Full density and potential representation}
The above-described ML density models are generally integrable into both pseudopotential-based and all-electron-based DFT methods. In this work, to combine with the full potential TB-MTO method,\cite{ZhangA2024,ZhangA2025} both $\rho_{DFT}$ and $\rho_{ref}$ are represented using a three–component scheme consisting of a smooth uniform–grid density for the interstitial region and atom–centered radial densities inside muffin–tin spheres, ensuring an exact reconstruction of the full density,
\begin{equation}\label{FullDensity}
   \rho(\vec{r})=n_{0}(\vec{r})+\sum_R[n_{1R}(\vec{r}_R)-\tilde{n}_{2R}(\vec{r}_R)]
\end{equation}
where $n_{0}$ denotes the smooth density on a uniform grid in the whole space, $n_{1R}$ denotes the real density inside the MT potential sphere, $\tilde{n}_{2R}$ is smooth and is introduced to cancel out the $n_{0}$ inside the MT spheres. Here, $n_{1R}$ and  $\tilde{n}_{2R}$ are represented on a set of spherical harmonics.
Accordingly, the full effective potential, including the Hartree and exchange-correlation potentials, can be expressed as
\begin{equation}\label{Potential}
   V_{eff}(\vec{r})= \tilde{V}^{0}(\vec{r})+\sum_{RL} [V_{RL}^{1}(\vec{r}_R)-\tilde{V}_{RL}^{2}(\vec{r}_R)]
\end{equation}
where $\tilde{V}^{0}(\vec{r})$ is the smooth-potential defined on the real-space mesh, $V_{RL}^{1}$ is the true muffin-tin potential inside the sphere at site $R$, and $\tilde{V}_{RL}^{2}$ subtracts the contribution of $\tilde{V}^{0}(\vec{r})$ within the spheres.
After obtaining $\Delta\rho_{DFT}$ for the case of all electron from DFT, the charge (of $\Delta \rho$) contained in the near-core region is small, but it features sharp high-frequency cusps that are numerically difficult to resolve on a coarse uniform grid. we thus apply a Gaussian smoothing within a cutoff radius around each atomic nucleus while preserving the total charge and multipole moments. The resulting smoothed density $\Delta\rho$ is therefore suitable as the learning target.
For compatibility with the downstream electronic-structure and transport codes, $\rho_{pred}$ is converted back into the three-component form as in Eq.\ref{FullDensity} used by the underlying full-potential TB-MTO framework for property calculations,e.g. band structure, quantum transport through the device structures.

\subsubsection{Fully relativistic TB-MTO method}

In this study, we carry out electronic-structure and quantum transport calculations employing the tight-binding muffin-tin orbital (TB-MTO) formulated within the kinked partial wave (KPW) framework introduced by Andersen and co-workers\cite{Andersen1995,Andersen2000,Andersen2000PRB,Vitos2007}. 
The KPWs constitute a minimal and strongly localized basis set and show important applications to quantum transport simulation of device structures.\cite{ZhangQ2019,Chen2020,ZhangY2021,Chen2023} Their strong spatial localization arises from the use of screened spherical waves (SSWs) as envelope functions, which effectively reduce intersite overlaps. This construction yields a basis that is both physically transparent and computationally efficient, making it particularly suitable for large-scale materials and device simulations.
 

For systems containing heavy elements, relativistic effects, especially spin–orbit coupling, are essential for an accurate description of the electronic structure. To account for these effects, we formulate the Hamiltonian within a fully relativistic framework, namely considering the Dirac equation
\begin{equation}\label{Dirac}
    [c\vec{\alpha}\cdot \vec{p}+(\beta-I_{4})c^2+V_{eff}(\vec{r})]\chi(\epsilon,\vec{r})=\epsilon \chi(\epsilon,\vec{r}),
 \end{equation}
where $c$ is the speed of light, 
$\vec{\alpha}=\begin{pmatrix}
0 & \vec{\sigma}\\
\vec{\sigma} & 0
\end{pmatrix}$ and 
$\beta=\begin{pmatrix}
I_2 & 0\\
0 & -I_2
\end{pmatrix}$, with $\vec{\sigma}$ denoting the Pauli matrices and $I_2$ the $2\times2$ identity matrix.
In the present work, we employ a double-$\kappa$ four-component KPWs basis set evaluated at two different energies. The relativitic KPW centered at site $R$ is defined as\cite{Pourovskii2005,Chen2020,Chen2023,Turek1997}
\begin{equation}\label{KPW}
   \Psi_{R\Lambda}(\epsilon,\vec{r})= \phi_{R\Lambda}(\epsilon,\vec{r})-\varphi_{R\Lambda}(\epsilon,\vec{r})+\psi_{R\Lambda}(k^2,\vec{r}),
\end{equation}
where the relativistic quantum numbers are denoted by $\Lambda=\kappa\mu$, the kinetic energy $k^2=\epsilon-V_{const}$ with $V_{const}$ a constant potential. $\phi_{R\Lambda}$ is the partial wave defined inside the potential sphere ($s_R$), $\psi_{R\Lambda}$ is the screened spherical wave (SSW) defined outside the screening sphere ($a_R$),and $\varphi_{R\Lambda}$ is the free electron solution defined in the interstitial region between the $s$- and $a$-spheres.  $\phi_{R\Lambda}$ and $\varphi_{R\Lambda}$ are connected continuously in first order at $s_R$, and  $\psi_{R\Lambda}$ and $\varphi_{R\Lambda}$ are connected continuously in zeroth order at $a_R$.\cite{Andersen1995,Andersen2000,Andersen2000PRB,Vitos2007}


\subsubsection{Full potential implementation}
To implement the full-potential formulation, we augment the KPW by adding a smooth function inside $a_R$ to restore the second-order continuity at the boundary of the screening sphere $a_R$.\cite{ZhangA2024,ZhangA2025} The resulting basis function is written as
\begin{equation}\label{KPW2}
   \Psi_{R\Lambda}(\vec{r})= \tilde{\Psi}_{R\Lambda}(\vec{r})+\sum_{R'\Lambda'} [\bar{\phi}_{R'\Lambda'}^{R\Lambda}(\vec{r}_{R'})-\tilde{\phi}_{R'\Lambda'}^{R\Lambda}(\vec{r}_{R'})]
\end{equation}
where $\bar{\phi}_{R'\Lambda'}^{R\Lambda}(\vec{r}_{R'})$ and $\tilde{\phi}_{R'\Lambda'}^{R\Lambda}(\vec{r}_{R'})$ denote the true KPW and smoothed wave inside the $a$-sphere, respectively. $\tilde{\Psi}_{R\Lambda}(\vec{r})$ is constructed such that it matches continuously with the true KPW at the screening spheres, given as
\begin{equation}\label{Smooth}
\tilde{\Psi}_{R\Lambda}(\vec{r})=\Psi_{R\Lambda}^{I}(\vec{r})+\sum_{R'\Lambda'}\tilde{\phi}_{R'\Lambda'}^{R\Lambda}(\vec{r}_{R'}),
\end{equation}
which $\Psi_{R\Lambda}^{I}(\vec{r})$ represents the KPW in the interstitial region. $\tilde{\Psi}_{R\Lambda}(\vec{r})$ is a short-ranged function and is rather smooth. Accordingly, two different grid systems are employed: a uniform and relatively coarse real-space grid for $\tilde{\Psi}_{R\Lambda}(\vec{r})$ and dense radial grids inside screening spheres. This double-grid strategy enables an accurate and computationally efficient calculation of the overlap and Hamiltonian matrices.\cite{ZhangA2025}

Based on Eq.\ref{KPW2}, the overlap and potential matrix elements naturally decompose into three components
\begin{equation}\label{Overlap}
\begin{split}
   O_{R'\Lambda',R\Lambda}
   &= \langle \tilde{\Psi}_{R'\Lambda'}|\tilde{\Psi}_{R\Lambda}\rangle \\\\
   &+\sum_{R_{1}\Lambda_{1}} [\langle \bar{\phi}_{R_{1}\Lambda_{1}}^{R'\Lambda'}| \bar{\phi}_{R_{1}\Lambda_{1}}^{R\Lambda}\rangle -\langle \tilde{\phi}_{R_{1}\Lambda_{1}}^{R'\Lambda'}| \tilde{\phi}_{R_{1}\Lambda_{1}}^{R\Lambda}\rangle];
\end{split}
\end{equation}
\begin{equation}\label{Potential2}
\begin{split}
   V_{R'\Lambda',R\Lambda}
= & \langle \tilde{\Psi}_{R'\Lambda'}|\tilde{V}^{0}|\tilde{\Psi}_{R\Lambda}\rangle+\sum_{R_{1}\Lambda_{1}\Lambda_{2}L_{3}}[\langle \bar{\phi}_{R_{1}\Lambda_{1}}^{R'\Lambda'} \\\\
& |V_{R_{1}L_{3}}^{1}|\bar{\phi}_{R_{1}\Lambda_{2}}^{R\Lambda}\rangle-\langle \tilde{\phi}_{R_{1}\Lambda_{1}}^{R'\Lambda'}|\tilde{V}_{R_{1}L_{3}}^{2}| \tilde{\phi}_{R_{1}\Lambda_{2}}^{R\Lambda}\rangle].
\end{split}
\end{equation} 
The first terms of Eqs. \ref{Overlap} and \ref{Potential2} are evaluated on a uniform real-space grid, while the others use radial grids within screening spheres. The third term cancels out the $a$-sphere contributions of the first term, leaving the interstitial contribution.

For the kinetic-energy matrix, a different strategy is adopted due to the triple-valued nature of the basis functions in the region between the $s$- and $a$-spheres and the presence of kinks at the boundaries of the $a$-spheres.
\begin{equation}\label{Kinetic}
\begin{split}
    T_{R'\Lambda',R\Lambda}&=-\phi_{R'\Lambda'}(a_{R'})K_{R'\Lambda',R\Lambda} + k^2\langle \Psi_{R'\Lambda'}|\Psi_{R\Lambda}\rangle \\\\
    &+\langle \Psi_{R'\Lambda'}|\epsilon-k^2-V_{0}(\vec{r}_R)|\phi_{R\Lambda}\rangle,
\end{split}
 \end{equation}
where $K_{R'\Lambda',R\Lambda}$ is the kink matrix.\cite{Andersen1995,Andersen2000,Andersen2000PRB,Vitos2007} Here, the first term arises from the kinks at the $a$-sphere boundaries, the second term is directly obtained from the overlap matrix, and the third term is evaluated via radial integrations inside the $s$-spheres. (See refs.\onlinecite{ZhangA2024,ZhangA2025} for more details.)

  \begin{table*}[htbp]
      \centering
      \label{tab:model_comparison}
      \caption{Comparison of the $\Delta \rho$ MSE (mean square error) and total charge deviation $\Delta Q$ (with the unit $e$) for three different models, including the primary model without charge-neutrality constraint, the primary model, and the full model with residual compensation.  The five Au/Hg$_{x}$Cd$_{1-x}$Te/Au structures are generated randomly with different $x$.}
      \renewcommand{\arraystretch}{1.5}
      \setlength{\tabcolsep}{8pt}
      \begin{tabular}{|l|c|c|c|c|c|c|}
         \hline
         \multirow{2}{*}{Structures} 
         & \multicolumn{2}{c|}{No-Conservation} 
         & \multicolumn{2}{c|}{Primary Model} 
         & \multicolumn{2}{c|}{Full Model} \\
         \cline{2-7}
         & MSE & $\Delta Q$ & MSE & $\Delta Q$ & MSE & $\Delta Q$ \\
         \hline
         Au$_{216}$/Hg$_{150}$Cd$_{250}$Te$_{416}$/Au$_{252}$ & 1.352e-8 & -0.186622 & 1.227e-8 & 0.076970 & 1.224e-8 & -0.005492 \\
         \hline
         Au$_{216}$/Hg$_{224}$Cd$_{176}$Te$_{416}$/Au$_{252}$ & 1.385e-8 & -0.061293 & 1.270e-8 & 0.062877 & 1.291e-8 & -0.005544 \\
         \hline
         Au$_{216}$/Hg$_{282}$Cd$_{118}$Te$_{416}$/Au$_{252}$ & 1.393e-8 & 0.050607 & 1.293e-8 & 0.068578 & 1.291e-8 & -0.006231 \\
         \hline
         Au$_{216}$/Hg$_{320}$Cd$_{80}$Te$_{416}$/Au$_{252}$ & 1.387e-8 & 0.059021 & 1.302e-8 & 0.055530 & 1.300e-8 & -0.006689 \\
         \hline
         Au$_{216}$/Hg$_{356}$Cd$_{44}$Te$_{416}$/Au$_{252}$ & 1.385e-8 & 0.158186 & 1.316e-8 & 0.056537 & 1.314e-8 & -0.007183 \\
         \hline
      \end{tabular}
      \label{tab:Delta_rho_MSE_comp_tab}
   \end{table*}

\subsection{Random-NEGF based quantum transport}

Once the overlap and Hamiltonian matrices are obtained for large-scale device structures, we employ the random nonequilibrium Green’s function (rNEGF) method\cite{TangM2024,ZhangQ2024} to compute electron transport through the device. Conventionally, the transmission for a two-probe device is given by\cite{Datta2005}
\begin{equation}\label{TransFormula}
T(E)=\mathrm{Tr}[G_{cc}^{\mathcal{R}} \Gamma_l G_{cc}^{\mathcal{A}} \Gamma_r].
\end{equation}
where $G_{cc}^{\mathcal{R}}$ and $G_{cc}^{\mathcal{A}}=G_{cc}^{\mathcal{R}\dag}$ are the respective retarded and advanced Green’s functions of the central region, $\Gamma_{l/r}$ are linewidth matrices for coupling to the electrodes.\cite{Datta2005} To achieve randomized sampling of the incident states, we perform an eigen-state decomposition of the Hermitian $\Gamma_l$ and introduce, 
\begin{equation}\label{SigmaLess1} \tilde{\Gamma}_{l}=U_{l}\sqrt{V_{l}}\tilde{I}_{l}\sqrt{V_{l}}U_{l}^{\dagger}.
\end{equation}
 Here, the random projector $\tilde{I}_{l}$ is defined as $\frac{1}{N_s}\sum_{j=1}^{N_s} \left| X_{j}\right\rangle \left\langle X_{j} \right|$,
where the random state $\left| X_{j} \right\rangle$ is a superposition of the $N_{mode}$ source incident eigen-modes, and is generated by randomly choosing 1 or -1 for each element.\cite{ZhangQ2024}
By construction, the ensemble average satisfies $\langle\tilde{I}_{l}\rangle=I_{l}$, ensuring the statistical correctness of the rNEGF approach. $\tilde{\Gamma}_{l}$ can then be written as
$\tilde{\Gamma}_{l}=\frac{1}{N_s}\sum_{j=1}^{N_s}  \left| R_{l,j}\right\rangle \left\langle R_{l,j} \right|$
where $\left| R_{l,j}\right\rangle = U_{l}\sqrt{V_{l}}\left| X_{l,j}\right\rangle $.
Therefore, the transmission is evaluated statistically as
\begin{equation}\label{TransFormula2}
    \tilde{T}(E)  = \frac{1}{N_s}\sum_{j=1}^{N_s} \left\langle \alpha_{l,j} \right| \Gamma_r  \left| \alpha_{l,j}\right\rangle,
\end{equation}
where $\left| \alpha_{l,j}\right\rangle = G_{cc}^{\mathcal{R}} \left| R_{l,j}\right\rangle$ are referred to as random scattering states, obtained by solving
\begin{equation}\label{linearEq}
  (E^{\dag}O_{cc} - H_{cc} - \Sigma_l^{\mathcal{R}} - \Sigma_r^{\mathcal{R}}) \left| \alpha_{l,j}\right\rangle = \left| R_{l,j}\right\rangle. 
\end{equation}
where $O_{cc}$ and $H_{cc}$ are the respective overlap and Hamiltonian matrices of the central device, and $\Sigma_{l/r}^{\mathcal{R}}$ are retarded self-energies of the left and right electrodes. The calculation of transmission is thereby reduced to the solution of \(N_s\) linear systems and $N_s$ is usually much smaller than $N_{mode}$ in a large device. In this work, we apply the direct solver based on LU factorization. 

\section{Results} \label{resultsec}

\subsection{ Training ML $\Delta\rho$ model }

All the systems in the training and test datasets are calculated by DFT with the full-potential LMTO method implemented in QUESTAAL.\cite{Pashov2020} Single-$\kappa$ bases, LDA exchange-correlation functional, and a G-vector cutoff of 9 are adopted for all calculations. For bulk cubic supercells Hg$_{32-n}$Cd$_n$Te$_{32}$, Hg$_{31-n}$Cd$_n$Te$_{32}$ and Hg$_{107-n}$Cd$_n$Te$_{108}$ with single Hg vacancy defects, 5$\times$5$\times$5 and 2$\times$2$\times$2 $k$-meshes are employed, respectively. For device structures, a 5$\times$5$\times$1 $k$-mesh is adopted. All DFT calculations are converged with a tolerance 10$^{-5}$ Ry in total energy. The experimental lattice constants of Au, HgTe, and CdTe are 7.71, 12.21, and 12.25 Bohr, respectively. A common lattice constant of 12.23 Bohr is used for all bulk and defect-containing Hg$_{1-x}$Cd$_x$Te alloys in this work. For the device structures, lattice matching in the plane perpendicular to the transport direction was achieved by commensurating a 3$\times$3 Au supercell with a 2$\times$2 Hg$_{1-x}$Cd$_x$Te supercell, resulting in an in-plane supercell lattice constant of 24.45 Bohr. Under this construction, the lattice constant of the Au electrodes corresponds to a 5.7\% tensile strain relative to the experimental bulk value. The transport direction of the device structures is aligned with the (100) crystal orientation of both Au and Hg$_{1-x}$Cd$_x$Te.

Training datasets for the primary model include (as shown in Fig.\ref{pred electron density fig1}):

(I) Bulk structures: 172 Hg$_{32-n}$Cd$_n$Te$_{32}$ with random configurations that cover different concentrations of Cd, together with bulk Au cubic supercells of 2 different sizes.

(II) Device structures: 400 Au$_{108}$/Hg$_{56-n}$Cd$_{n}$Te$_{64}$/- Au$_{126}$ structures with random configurations that cover different concentrations of Cd in the channel. On the interface Au/Te, atomic structure is relaxed with VASP.

(III) Defective bulk structures: 10 Hg$_{31-n}$Cd$_n$Te$_{32}$ and 12 Hg$_{107-n}$Cd$_n$Te$_{108}$ supercells with different concentrations of Cd. The atomic structure around the $Hg/Cd$ vacancy is relaxed with VASP.\cite{Kresse1999} 

The dataset for training the residual compensation model includes the following (as shown in Fig.\ref{pred electron density fig1}): 

(IV) the bulk structures of (I) (III);

(V) 50 Au$_{108}$/Hg$_{152-n}$Cd$_{n}$Te$_{160}$/Au$_{126}$ structures with different concentrations of Cd, where the increased channel length prevents the coupling between the left and right interfaces.

 \begin{figure*}[htbp]
     \centering
     \includegraphics[width=0.9\linewidth]{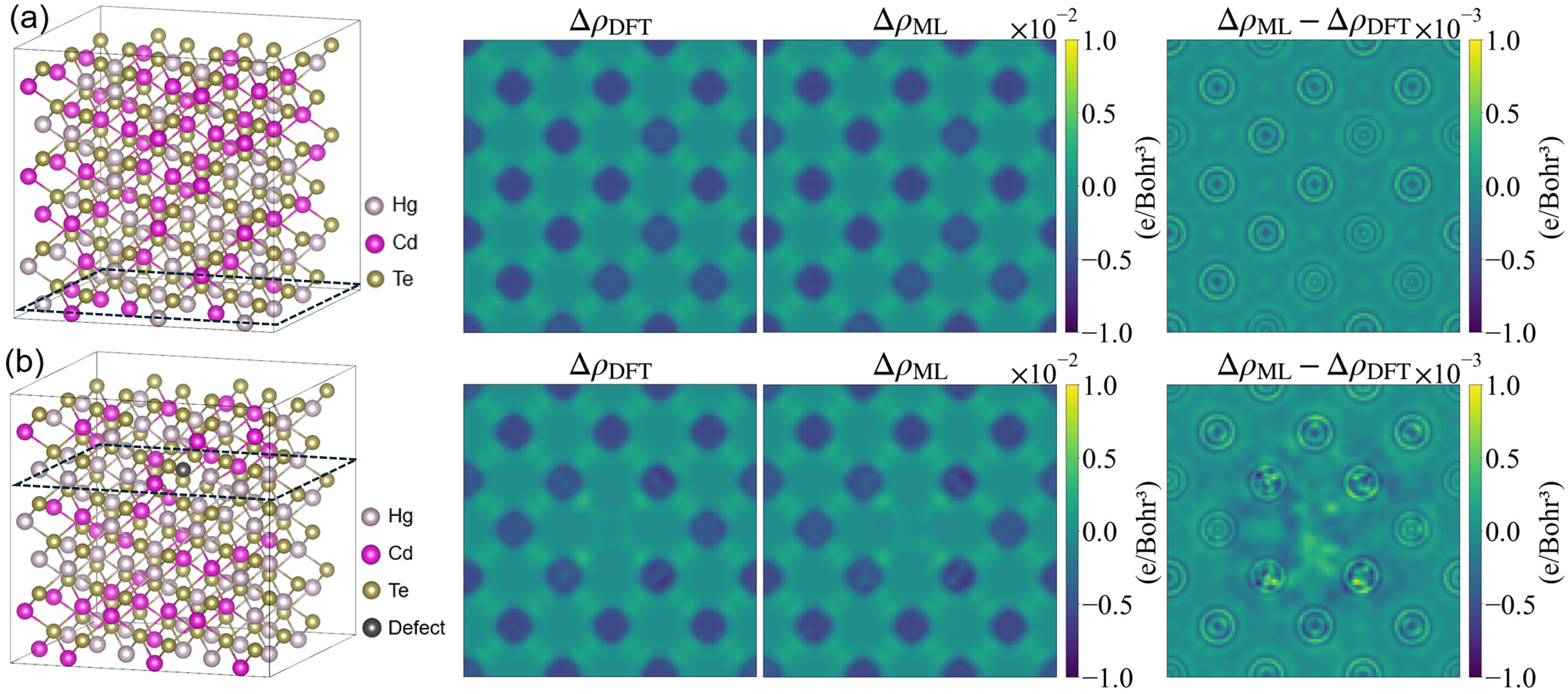}
    \caption{Comparison of ML $\Delta \rho$ with fully self-consistent DFT result for two bulk systems, including (a) for Hg$_{54}$Cd$_{54}$Te$_{108}$ supercell, (b) for Hg$_{67}$Cd$_{40}$Te$_{108}$ supercell with a single Hg vacancy. For each system, we show the random atomic strutures, 2D slices for $\Delta\rho_{DFT}$, $\Delta\rho_{ML}$, and the density loss $\Delta\rho_{ML}-\Delta\rho_{DFT}$.}
    \label{pred-bulk-density}
 \end{figure*}

In the rest of this paper, all the tested bulk and device structures containing Hg$_{1-x}$Cd$_{x}$Te are generated randomly, without correlation to the training dataset.
 We first investigate the MSE and total charge error $\Delta Q$ with three different models for 5 random device structures $Au_{216}$/Hg$_{400-n}$Cd$_n$Te$_{416}$/Au$_{252}$ in the Table.\ref{tab:Delta_rho_MSE_comp_tab}.  All three models, including the primary model without the charge conservation constraint, the primary model, and the full model, give nearly identical MSE values around $1.3\times10^{-8}$ for all tested device structures. However, compared with the primary model without the charge conservation constraint, both the primary model and the full model achieve markedly improved and more stable global charge neutrality, with more than an order of magnitude reduction in $\Delta Q$. Taking the Au$_{216}$/Hg$_{150}$Cd$_{250}$Te$_{416}$/Au$_{252}$ system as an example, the unconstrained primary model yields a $\Delta Q$ of $-0.186622$, whereas the implementation of the conservation constraint reduces the magnitude to $0.076970$ with the primary model and $-0.005492$ with the full model. It should be noted that the residual compensation used in the full model further enhances local charge consistency without causing any noticeable degradation in either MSE or total-charge deviation, as will be seen later in the improved device potential profiles.

\subsection{Bulk structures of Hg$_{1-x}$Cd$_x$Te alloys}

As a basic validation of the ML density model, combined with the first-principles FP-TB-MTO approach, we investigate the electronic structures of HgCdTe random alloys in both defect-free and defective configurations. The atomic structures, $\Delta\rho_{DFT}$, $\Delta\rho_{pred}$, and the density loss $\Delta\rho_{DFT}-\Delta\rho_{pred}$ for the cubic supercells Hg$_{54}$Cd$_{54}$Te$_{108}$ and Hg$_{67}$Cd$_{40}$Te$_{108}$ with a single Cd/Hg vacancy are presented in the respective Fig.~\ref{pred-bulk-density} (a) and (b). It is observed that, for both structures, the $\Delta \rho_{pred}$ 2D slices are in close agreement with the DFT results in both spatial distribution and amplitude. The minor discrepancies in the density loss are largely confined to the near-core region and the vicinity of the defect. The absolute density error of all grid points is smaller than $1\times10^{-3}$ $e/Bohr^{3}$, suggesting that the model can reliably capture the essential charge redistribution not only in compositionally disordered systems but also in defect-induced locally distorted environments.

\begin{figure*}[htbp]
    \centering
    \includegraphics[width=0.9\linewidth]{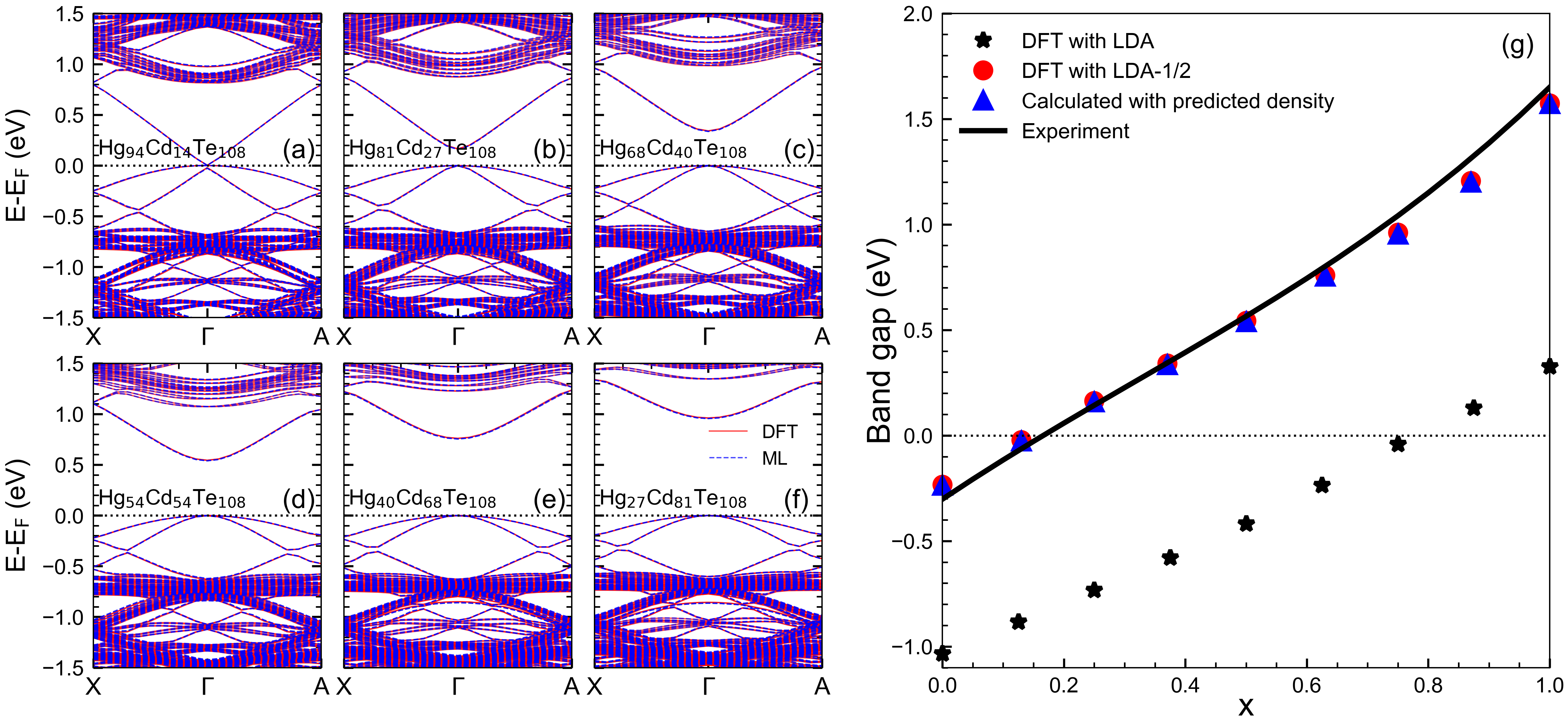}
  \caption{Electronic band structures and composition-dependent band gaps of Hg$_{1-x}$Cd$_x$Te alloys. (a)–(f) show the ML-density-based (in dashed) band structures of Hg$_{1-x}$Cd$_x$Te compared to DFT results (in solid) for $x$=0.13, 0.25, 0.37, 0.50, 0.63, and 0.75, respectively. (g) presents the LDA-1/2 band-gap results for the DFT density (red dots) and ML density (triangles), compared to LDA results(asterisks) and experimental data (solid line by $E_g(x)=-0.302+1.93x-0.81x^2+0.832x^3$).\cite{Hansen1982} } \label{BandsHgCdTe}
 \end{figure*}

As shown in Fig.~\ref{BandsHgCdTe} (a)-(f), the band structures of Hg$_{1-x}$Cd$_x$Te alloys for $x$=0.13, 0.25, 0.37, 0.5, 0.63, 0.75 calculated for the cubic supercells Hg$_{108-n}$Cd$_{n}$Te$_{108}$ using the ML densities are compared with the results of DFT. To construct the tight-binding Hamiltonians for these systems, a real-space cutoff radius of 6.1 Bohr is employed for SSWs in FP-TB-MTO. The radii of the screening spheres for Hg, Cd, and Te atoms are set to 2.39 Bohr. The LDA-1/2 is incorporated to correct the band gap for all HgCdTe alloys.\cite{Ferreira2008,Ferreira2011} The band structures derived from the predicted charge densities (dashed lines) exhibit very good agreement with the DFT results (solid lines) across all the concentrations of Cd, presenting the deviations below 10 meV. This remarkable agreement demonstrates that the ML densities capture the essential features of the electronic ground state with high precision, even for complex alloy configurations.

Fig.\ref{BandsHgCdTe} (g) further presents the evolution of the band gap as a function of Cd concentration $x$. The theoretical band gaps were corrected using the LDA-1/2 scheme and are plotted as discrete symbols, while the experimental data are represented by the solid black line. The calculated band gap values reproduce the experimental trend well over the entire composition range, including the transition from the semimetallic regime at low Cd content to the widening band gap in Cd-rich alloys. In particular, the transition occurs around $x$=0.166, where the system undergoes a fundamental topological phase transition, characterized by the band inversion between the $s$-like $\Gamma_6$ band and the $p$-like $\Gamma_8$ quadruplet. In the vicinity of this transition point, the band structure near the Fermi level manifests as a unique superposition: the interaction between the $\Gamma_6$ and light-hole $\Gamma_8$ bands creates a linear, Dirac-like dispersion of massless fermions, while the heavy-hole $\Gamma_8$ band remains parabolic and pinned at the $\Gamma$ point, as shown in Fig. \ref{BandsHgCdTe} (a). This coexistence of linear and parabolic bands at the critical point is a hallmark of the Kane model\cite{Kane1957} physics that governs the exceptional carrier mobility in these materials.
The consistency between the theoretical and experimental results confirms the reliability of combining ML densities with the LDA-1/2 approach for band-gap estimation. For comparison, the band gaps calculated using LDA are also plotted and denoted by asterisks which presents large underestimation compared to experimental values. Overall, these results indicate that the proposed ML-assisted framework provides an efficient and accurate pathway for predicting the electronic structures of alloyed semiconductors. The good agreement with both DFT and experimental benchmarks highlights its potential as a tool for large-scale first-principles simulations of semiconducting materials and devices.

 \begin{figure}[htbp]
    \centering
    \includegraphics[width=0.8\columnwidth]{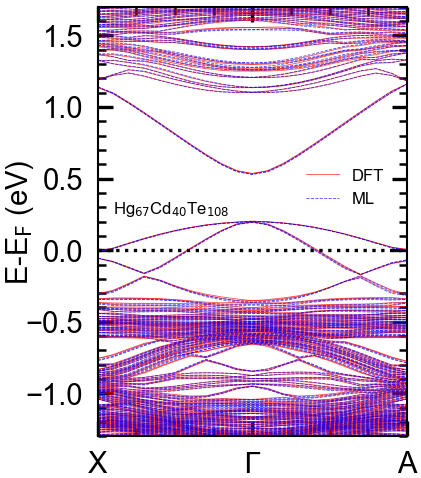} 
    \caption{Band structure of Hg$_{67}$Cd$_{40}$Te$_{108}$ supercell with a single Cd/Hg vacancy (as shown in Fig.\ref{pred-bulk-density}(b)).}
    \label{Bands_HgVac}
\end{figure}

 \begin{figure*}[htbp]
    \centering
    \includegraphics[width=0.9\linewidth]{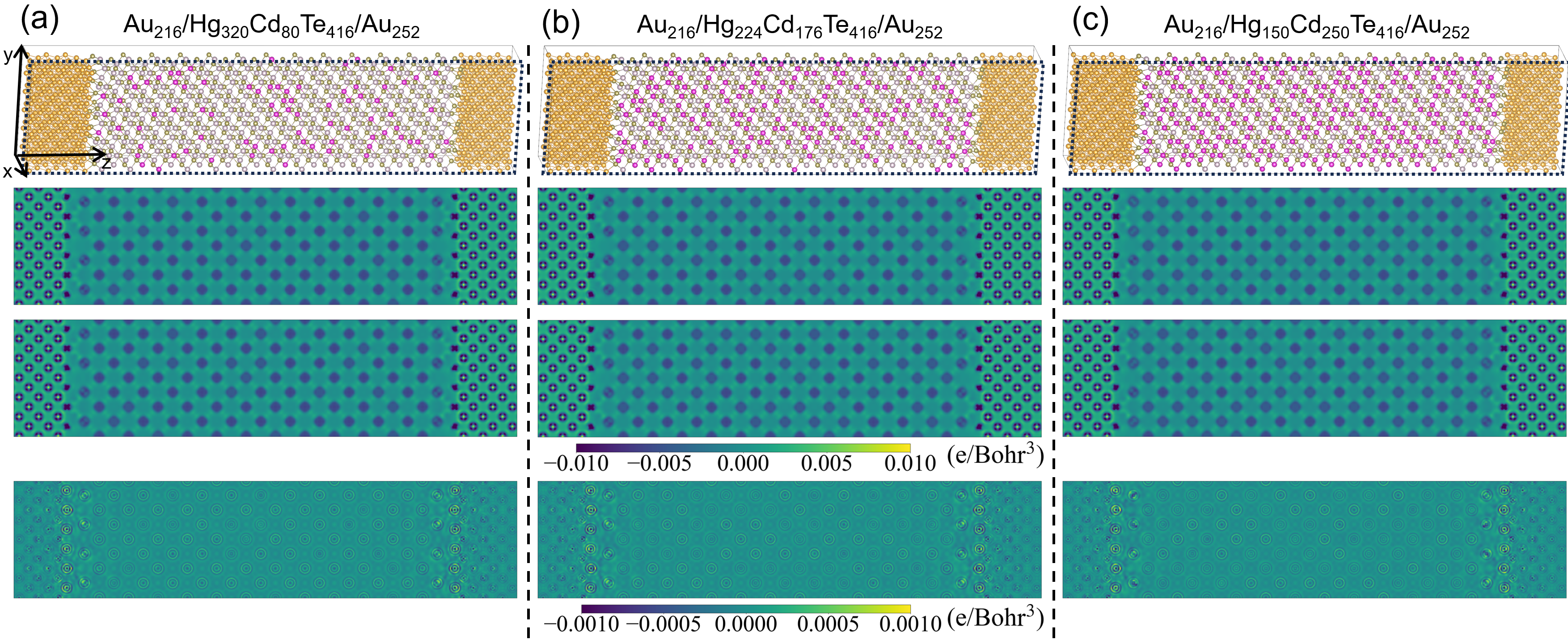}
     \caption{Comparison of atomic structures, $\Delta\rho_{DFT}$, $\Delta\rho_{ML}$, and the corresponding error (Loss Density) for three device configurations: (a) Au$_{216}$/Hg$_{320}$Cd$_{80}$Te$_{416}$/Au$_{252}$, (b) Au$_{216}$/Hg$_{224}$Cd$_{176}$Te$_{416}$/Au$_{252}$, and (c) Au$_{216}$/Hg$_{150}$Cd$_{250}$Te$_{416}$/Au$_{252}$. All the device structures are periodic in the x-y plane.
     For each case, panels from top to bottom display the random atomic structure, 2D slices of $\Delta\rho_{DFT}$ and $\Delta\rho_{ML}$, and the density loss $\Delta\rho_{ML}-\Delta\rho_{DFT}$.}
     \label{ElectronDensityDevice}
 \end{figure*}
 
Fig.\ref{Bands_HgVac} presents the electronic band structure of the HgCdTe system containing a single Cd/Hg vacancy as shown in Fig.\ref{pred-bulk-density}(b), calculated using DFT (solid) and ML densities (dashed),   respectively. Overall, the two sets of band structures show an excellent level of agreement over the entire energy window and along the full high-symmetry path. Both the dispersion relations and the relative ordering of the bands are faithfully reproduced by the ML-based calculation. In particular, the defect-induced states and their hybridization with the host bands are captured with high accuracy, indicating that the ML-predicted charge density correctly reflects the local electronic reconstruction around the Cd/Hg vacancy.

Near the band edges, which are most relevant for transport and optical properties, the ML-derived bands closely track the DFT results, with no noticeable artificial gap opening, band crossing errors, or spurious states. This agreement demonstrates that the ML model preserves not only the global electrostatic potential but also its subtle spatial variations induced by the defect, which are crucial for describing localized or resonant defect states. At higher energies, where the band density becomes large, the ML and DFT bands remain nearly indistinguishable, further confirming the robustness of the approach. The agreement between the two calculations demonstrates that the ML charge density provides a reliable approximation to the fully self-consistent DFT charge density, even in the presence of point defects that induce strong local perturbations. This result is particularly significant for simulation of large-scale realistic materials and devices, where fully self-consistent DFT calculations become prohibitively expensive. 

\subsection{Device structures of Au/Hg$_{1-x}$Cd$_x$Te/Au}

To further verify the accuracy of the ML model, we investigate the electronic structure, potential energy profile and quantum transport properties of two-probe devices consisting of a disordered Hg$_x$Cd$_{1-x}$Te region sandwiched by semi-infinite left and right Au electrodes. Fig.\ref{ElectronDensityDevice}  (a)-(c) present the random atomic structures, $\Delta\rho_{DFT}$, $\Delta\rho_{ML}$, and the density loss for the respective Au$_{216}$/Hg$_{320}$Cd$_{80}$Te$_{416}$/Au$_{252}$, Au$_{216}$/Hg$_{224}$Cd$_{176}$Te$_{416}$/Au$_{252}$, and Au$_{216}$/Hg$_{150}$Cd$_{250}$ -Te$_{416}$/Au$_{252}$.
It is observed that, for all three device structures, $\Delta \rho_{pred}$ are in close agreement with the DFT results in both spatial distribution and magnitude. It is clear that the minor density losses are mainly localized at the interfaces. The absolute charge-density error remains below $1\times10^{-3}$ $e/Bohr^{3}$, indicating that the ML model reliably captures the essential features of charge redistribution in the device structures.

\begin{figure*}[htbp]
\centering
\includegraphics[width=0.9\linewidth]{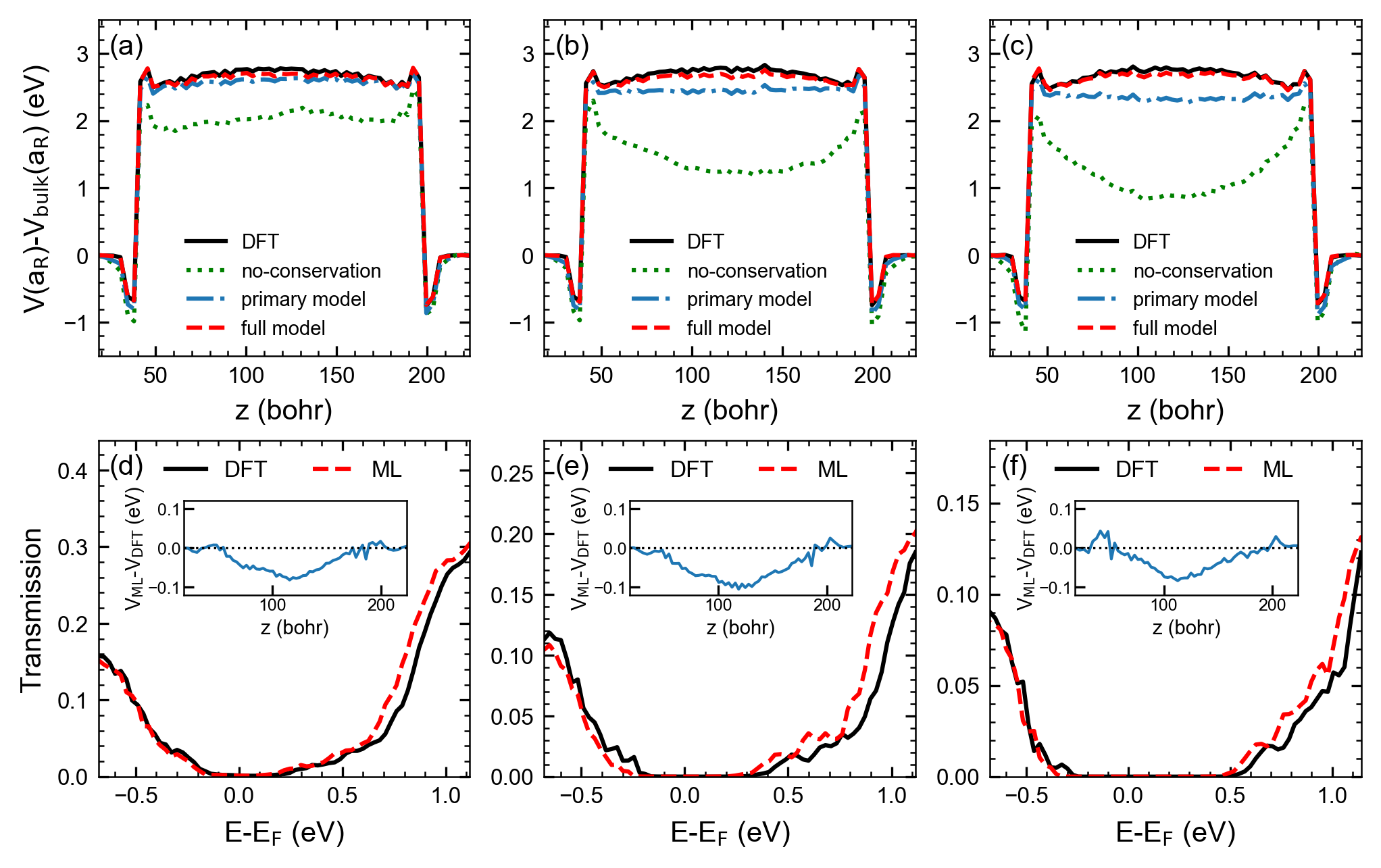}
\caption{Comparison of DFT- and ML-based potential energy and quantum-transport properties for three Au/Hg$_{1-x}$Cd$_x$Te/Au random device structures, including (a,d) for Au$_{216}$/Hg$_{320}$Cd$_{80}$Te$_{416}$/Au$_{252}$, (b,e) for Au$_{216}$/Hg$_{224}$Cd$_{176}$Te$_{416}$/Au$_{252}$, and (c,f) for Au$_{216}$/Hg$_{150}$Cd$_{250}$Te$_{416}$/Au$_{252}$. (a)–(c) show the layer-resolved potential energy difference, $V(a_R)-V_{bulk}(a_R)$, evaluated at the MT sphere radius of a selected atom in each atomic layer, for the ML model without charge-neutrality constraint ("no-conservation"), primary and full models, and compare with the result of fully self-consistent DFT. (d)–(f) show the transmission per unit area (66.44 Bohr$^2$) versus energy calculated from Hamiltonians constructed using the DFT and  ML densities. The insets in (d)–(f) show the potential energy deviation, $V_{ML}-V_{DFT}$, corresponding to (a)-(c).
}\label{Pot_Trans}
\end{figure*}

Figure \ref{Pot_Trans} (a)–(c) show the layer-resolved potential energy difference for the three device structures in Fig. \ref{ElectronDensityDevice} (a)–(c), defined as the difference between the potential (including Hartree and exchange-correlation potentials) at the screening sphere ($a_R$) of an atom in each layer and its bulk reference. For all device calculations, $a_R$ for Au and interfacial Au atoms are set to 2.58 and 2.40 Bohr, respectively.  The Au bulk potential serves as the boundary condition (0.0 eV in the figure). This difference directly reflects modifications to the local potential across the device due to charge redistribution near metal/semiconductor interfaces.

In all panels, fully self-consistent DFT results (black solid curves) serve as the benchmark. It is observed that the potential energy in the metallic Au buffer layers quickly approaches the bulk Au value within a few atomic layers away from the Au/Hg$_{1-x}$Cd$x$Te interface, demonstrating the significant screening effects of metals. In contrast, a pronounced variation in the local potential emerges around the interfaces. For example, the potential of two adjacent Au layers at the Au/Hg$_{1-x}$Cd$_x$Te interface decreases from 0 to approximately –0.65 eV and –0.70 eV, respectively, while the potential of the Te atom at the interface increases by about 2.5 eV relative to its corresponding bulk value. This behavior originates from interfacial charge transfer: electrons flow from the Au electrode toward the Hg$_{1-x}$Cd$_x$Te, giving rise to a strong built-in electric field confined near the interface and thus resulting in a significant potential drop on the Au side and potential increase within the Hg$_{1-x}$Cd$_x$Te. This redistribution of charge effectively forms interfacial dipoles at the two interfaces, and thus presents a long-range electrostatic potential, giving rise to the almost constant but large potential shift deep within Hg$_{1-x}$Cd$_x$Te as seen in all device structures. It is clear that HgCdTe alloys present much weaker screening effects, and take about 50 Bohr to fully screen the influence of the interface, and finally present a potential-energy plateau in the middle of the device structures, indicating the recovery of the bulk-like electronic structure in the central HgCdTe alloys.

As shown for all structures, the green-dotted curves, obtained from the ML density without charge neutrality constraint (“no-conservation”), deviate dramatically from the DFT profiles. For instance, the potential deviations identified in the vicinity of the device center (located at \(z = 118.93\) Bohr) are 0.68, 1.51, and 1.88 eV for the respective device structures in Fig.\ref{Pot_Trans}(a)–(c). Not only is the overall magnitude incorrect, but the spatial variation across the device is also qualitatively incorrect. This demonstrates that omitting the charge neutrality constraint leads to substantial errors in the charge density everywhere, and consequently, to a severely distorted potential throughout the entire device structure.

 However, the primary model (blue-dash-dotted curves) significantly improves the potential differences approaching the DFT results and presents high stability for different device structures, highlighting the importance of enforcing charge neutrality in the ML density model. The primary model greatly improves the local charge neutrality inside the Hg$_{1-x}$Cd$x$Te region in all structures, giving rise to the potential plateaus seen in Fig. \ref{Pot_Trans}(a)–(c). At the interfaces and Au buffer layers, the primary model also reproduces the DFT results very well. However, despite the correct plateau behavior, non-negligible discrepancies from DFT persist inside the central Hg$_{1-x}$Cd$_x$Te region. The maximum deviations in the central part amount to 0.16, 0.33, and 0.46 eV in panels (a), (b), and (c), respectively. We attribute this residual error to the short channel length of Hg$_{1-x}$Cd$_x$Te in the device datasets used for training the primary model, which underestimates the screening effect of HgCdTe alloys. These observations indicate that, although the imposition of charge neutrality corrects the overall behavior and enhances numerical stability, local inaccuracies in the ML density still accumulate in the device interior and give rise to a residual potential mismatch.

By contrast, the full model (red dashed curves), which combines ML-predicted charge densities with both charge-neutrality enforcement and residual correction, agrees closely with the DFT results across the entire device for all three structures. The interface-induced potential steps, the central plateau, and the layer-by-layer variations are reproduced with a maximum deviation of about 0.1eV. This level of agreement indicates that the combined corrections substantially reduce both the global long-wavelength bias and the local residual errors in the ML charge density, yielding potential profiles that are quantitatively consistent with fully self-consistent DFT. The insets of Fig. \ref{Pot_Trans} (d)–(f) illustrate the deviations of the potential energies obtained from DFT and the full-model ML densities. Overall, for all three device structures, the maximal deviation remains within 0.1eV, indicating the high quality of the ML-predicted charge redistribution around the interfaces and local charge neutrality away from them. Across all three devices, the deviations in the leftmost and rightmost four Au layers are within 12 meV, consistently smaller than those near the Au/Hg$_{1-x}$Cd$_x$Te interfaces and within the central device region. This reflects the relatively uniform and strongly screened electronic environment in metallic Au, which is more readily captured by the ML model, whereas the interface and semiconductor regions involve more complex charge redistribution due to the random alloy configurations. The relatively larger errors in the Hg$_{1-x}$Cd$_x$Te region are attributed to the limited number and short channel length of the device structures in the training dataset.

\begin{figure*}[htbp]
\centering
\includegraphics[width=0.9\linewidth]{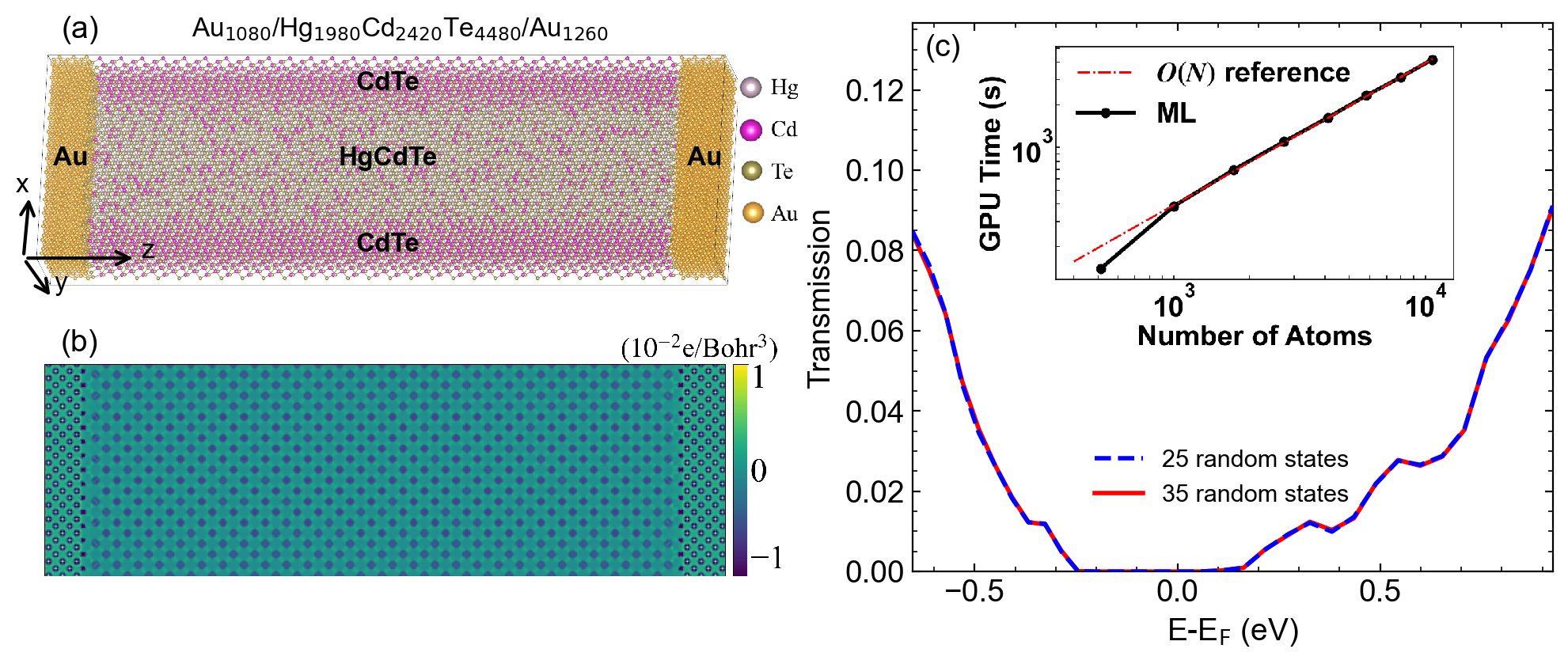}
\caption{
(a) Atomic structure of an infared photodetector device Au/[CdTe/Hg$_{1-x}$Cd$_x$Te/CdTe]/Au containing 11,220 atoms. Periodic boundary conditions are imposed in the $x$-$y$ plane. The central scattering region consists of a layered CdTe/Hg$_{1-x}$Cd$_x$Te superlattice oriented along $x$-direction. The Hg$_{1-x}$Cd$_x$Te alloy region is defined as the channel. (b) Spatial distribution of ML $\Delta \rho$ of a 2D slice in the x-y plane. (c) Transmission per unit area (66.44 Bohr$^2$) versus energy calculated on the basis of the ML density. The inset shows the linear scaling of GPU inference time with atom number.}
\label{device_large}
\end{figure*}

Fig.\ref{Pot_Trans} (d)-(f) presents a comparison of the electronic transmission (per unit area (66.44 Bohr$^2$)) obtained from the device Hamiltonian constructed using DFT and  ML densities.
Transport calculations were carried out using a $10\times10$ $k$-mesh in the 2D Brillouin zone of the $x$–$y$ plane. Here, due to the small number of incident states, transmission is explicitly calculated for all incident eigenmodes rather than using random sampling. 
Across the three compositions, the ML-based transmission reproduces the DFT result closely, with a small rigid energy shift of below 0.1eV.The systematic trend from (d) to (f) is consistently captured: as the Cd content increases, the band gap of the central region widens, and the energy window around the Fermi level in which transmission is suppressed (effectively zero) becomes progressively broader. This agreement, including the composition-dependent evolution of the transport gap, demonstrates that the ML-predicted charge densities capture the quantitative transport characteristics of the device, with the residual energy shift reflecting the small potential offset discussed above.

Fig.\ref{device_large} demonstrates the applicability of the ML density to quantum-transport calculations of a large device structure. As shown in Fig.\ref{device_large} (a), the device cell, periodic in x and y direction, comprises 11220 atoms and includes Au contacts, and disordered CdTe/Hg$_{1-x}$Cd$_x$Te/CdTe sandwich, posing a severe challenge for self-consistent DFT simulations. The ML-predicted $\Delta \rho$ in Fig.\ref{device_large} (b) exhibits quite accurate spatial distribution across the entire device, as compared to the $\Delta \rho$ results  in Fig.\ref{ElectronDensityDevice} and Fig.\ref{pred-bulk-density} near the interface Au/Hg$_{1-x}$Cd$_x$Te and in the interior region, demonstrating the high fidelity and robustness of the DeltaChargeNet. 
Fig.\ref{device_large}(c) presents the transmission spectrum (per unit area (66.44 Bohr$^2$)) computed from the ML density derived Hamiltonian based FP-TB-MTO method. A $2\times5$ $k$-mesh in the 2D Brillouin zone is adopted. Given the high density of incident states in the Au electrode (about 300 incident eigen-modes per $k$-point) in this large system, we utilized the rNEGF method with 35 random incident states  (red-solid line) for each $k$-point to ensure a low statistical error in transmission (compared to the results of 25 random states in blue-dashed line).\cite{ZhangQ2024} The overall energy dependence closely resembles that of the smaller reference device: the transmission decreases as the energy approaches the Fermi level from below, vanishes within an energy window of approximately 0.25 eV around the Fermi energy, and then increases monotonically at higher energies. The presence of a well-defined transport gap near the Fermi level is consistent with the underlying electronic structure of the Hg$_{1-x}$Cd$_x$Te channel and confirms that the ML density preserves the essential features governing quantum transport.
Taken together, these results demonstrate that ML density model can enable accurate potential energy alignment and reliable transport calculations in device structures far beyond the size accessible to fully self-consistent first-principles methods, while maintaining physically correct interface and bulk behavior.

Finally, we investigate the computational scalability of DeltaChargeNet for large-scale systems in the inset of Fig.~\ref{device_large}(c). The benchmark is carried out on a set of Hg$_{0.5}$Cd$_{0.5}$Te cubic supercells ranging from 512 to 10648 atoms. All inference measurements are performed on an NVIDIA RTX 3090 Ti GPU (24 GB), paired with an AMD 9950X CPU(4.3 GHz). The residual compensation model is excluded from the timing statistics because its cost is negligible relative to that of the primary model. As shown, for a structure with $10648$ atoms, DeltaChargeNet requires 4160 seconds inference time for a single prediction. Most importantly, the computational cost of the DeltaChargeNet scales linearly with the number of atoms, presenting the high scalability of the proposed framework.

\section{CONCLUSIONS}\label{conclusion}

In summary, we have developed an efficient and scalable machine-learning framework for predicting charge densities called DeltaChargeNet, and integrated it with the full-potential tight-binding muffin-tin orbital (FP-TB-MTO) method and non-equilibrium Green's function (NEGF) formalism to enable large-scale electronic structure and quantum transport simulations of materials and nanoelectronic devices. Our approach directly predicts electron densities from atomic configurations with low linear-scaling cost, bypassing the need for self-consistent field (SCF) iterations, while the FP-TB-MTO method provides an accurate tight-binding representation of the system Hamiltonian.
Using Hg$_{1-x}$Cd$_x$Te alloys and Au/Hg$_{1-x}$Cd$_x$Te/Au device structures as representative examples, we demonstrate that the trained machine-learning models can reproduce high-accuracy ground-state charge densities. Besides the high quality in reproducing the band structure of bulk Hg$_x$Cd$_{1-x}$Te, our method achieves first-principles accuracy in simulating realistic device structures, including potential energy profiles and quantum transport properties, as confirmed by the good agreement with full DFT SCF calculations.
The DeltaChargeNet accurately describes the interfacial dipole and preserves local charge neutrality in device structures.
Compared with KS-DFT approaches, our method substantially reduces computational cost while maintaining high accuracy for large-scale device simulations and materials simulations. By combining first-principles accuracy with machine-learning scalability, this work provides a practical route toward high-throughput electronic structure and quantum transport calculations for complex material systems and realistic electronic devices.

\begin{acknowledgments}
Y.K. and W.L. acknowledge financial support from NSFC (grant No.12227901). J.Y. acknowledges the NSFC (Grant No.12504288), the Science and Technology Commission of Shanghai Municipality (Grant No. 25ZR1402550). X.Z. acknowledges the robotic AI-Scientist platform of Chinese Academy of Sciences.
The authors thank the HPC platform of ShanghaiTech University for providing the computational facility.
\end{acknowledgments}


\begin{thebibliography}{99}


\bibitem{zhanglinfeng} 
L. Zhang, J. Han, H. Wang, R. Car, and W. E, Deep potential molecular dynamics: a scalable model with the accuracy of quantum mechanics. Phys. Rev. Lett. 120, 143001 (2018).

\bibitem{Behler2007} 
J. Behler and M. Parrinello, Generalized neural-network representation of high-dimensional potential-energy surfaces, Phys. Rev. Lett. 98, 146401 (2007).

\bibitem{Wang2018} 
H. Wang, L. Zhang, J. Han, and W. E. DeePMD-kit: A deep learning package for many-body potential energy representation and molecular dynamics, Comput. Phys. Commun 228, 178-184 (2018). 

\bibitem{Batzner2022} 
S. Batzner, A. Musaelian, L. Sun, M. Geiger, J. P. Mailoa, M. Kornbluth, N. Molinari, T. E. Smidt, and B. Kozinsky, E(3)-equivariant graph neural networks for data-efficient and accurate interatomic potentials. Nat. Commun 13, 2453 (2022).

\bibitem{Batatia2022} 
I. Batatia, D. P. Kovacs, G. Simm, C. Ortner, G. Csanyi, MACE: Higher order equivariant message passing neural networks for fast and accurate force fields, NIPS, 35, 11423-11436 (2022).

\bibitem{Zhang2024} 
D. Zhang, X. Liu, X. Zhang, C. Zhang, C. Cai, H. Bi, Y. Du, X. Qin, A. Peng, J. Huang, B. Li, Y. Shan, J. Zeng, Y. Zhang, S. Liu, Y. Li, J. Chang, X. Wang, S. Zhou, J. Liu, X. Luo, Z. Wang, W. Jiang, J. Wu, Y. Yang, J. Yang, M. Yang, F.-Q. Gong, L. Zhang, M. Shi, F.-Z. Dai, D. M. York, S. Liu, T. Zhu, Z. Zhong, J. Lv, J. Cheng, W. Jia, M. Chen, G. Ke, W. E, L. Zhang, and H. Wang, DPA-2: a large atomic model as a multi-task learner, npj Comput. Mater. 10, 293 (2024).

\bibitem{Li2022} 
H. Li, Z. Wang, N. Zou, M. Ye, R. Xu, X. Gong, W. Duan, and Y. Xu, Deep-learning density functional theory Hamiltonian for efficient ab initio electronic-structure calculation, Nat. Comput. Sci. 2, 367-377 (2022).

\bibitem{Zhang2022} 
L. Zhang, B. Onat, G. Dusson, A. McSloy, G. Anand, R. J. Maurer, C. Ortner, and J. R. Kermode, Equivariant analytical mapping of first principles Hamiltonians to accurate and transferable materials models, npj Comput. Mater. 8, 158 (2022). 

\bibitem{Zhong2023} 
Y. Zhong, H. Yu, M. Su, X. Gong, and H. Xiang, Transferable equivariant graph neural networks for the Hamiltonians of molecules and solids, npj Comput. Mater. 9, 182 (2023). 

\bibitem{Schutt2019} 
K. T. Schutt, M. Gastegger, A. Tkatchenko, K.-R. Müller, and R. J. Maurer, Unifying machine learning and quantum chemistry with a deep neural network for molecular wavefunctions, Nat. Commun. 10, 5024 (2019).

\bibitem{Gu2024} 
Q. Gu, Z. Zhouyin, S. K. Pandey, P. Zhang, L. Zhang, and W. E, Deep learning tight-binding approach for large-scale electronic simulations at finite temperatures with ab initio accuracy, Nat. Commun. 15, 6772 (2024). 

\bibitem{Chen2019} 
C. Chen, W. Ye, Y. Zuo, C. Zheng, and S. P. Ong, Graph networks as a universal machine learning framework for molecules and crystals, Chem. Mater. 31, 3564-3572 (2019).

\bibitem{Choudhary2021} 
K. Choudhary and B. DeCost, Atomistic Line Graph Neural Network for improved materials property predictions, npj Comput. Mater. 7, 185 (2021).
   
\bibitem{Xie2018} 
T. Xie and J. C. Grossman, Crystal graph convolutional neural networks for an accurate and interpretable prediction of material properties, Phys. Rev. Lett. 120, 145301 (2018).

\bibitem{Kosmala2023}
A. Kosmala, J. Gasteiger, N. Gao, and S. Günnemann, Ewald-based long-range message passing for molecular graphs, in Int. Conf. Mach. Learn. (PMLR, 2023).

\bibitem{Zhong2026} 
Y. Zhong, X. Li, X. Gong, and H. Xiang, Physics-Informed Long-Range Coulomb Correction for Machine-learning Hamiltonians, arXiv:2603.20007 (2026).

\bibitem{Zou2025} 
J. Zou, Z. Zhouyin, D. Lin, Y. Huang, L. Zhang, S. Hou, and Q. Gu, Deep learning accelerated quantum transport simulations in nanoelectronics: from break junctions to field-effect transistors, npj Comput. Mater. 11, 375 (2025).

\bibitem{Monti2012}
O. L. A. Monti, Understanding interfacial electronic structure and charge transfer: an electrostatic perspective, J. Phys. Chem. Lett. 3, 2342-2351 (2012).

\bibitem{Hohenberg1964}
P. Hohenberg and W. Kohn, Inhomogeneous electron gas, Phys. Rev. 136, B864 (1964).


\bibitem{Grisafi2018} 
A. Grisafi, A. Fabrizio, B. Meyer, D. M. Wilkins, C. Corminboeuf, and M. Ceriotti, Transferable machine-learning model of the electron density, ACS Cent. Sci. 5, 57-64 (2018).

\bibitem{Fabrizio2019} 
A. Fabrizio, A. Grisafi, B. Meyer, M. Ceriotti, and C. Corminboeuf, Electron density learning of non-covalent systems, Chem. Sci. 10, 9424-9432 (2019).

\bibitem{Dick2020} 
S. Dick and M. Fernandez-Serra, Machine learning accurate exchange and correlation functionals of the electronic density, Nat. Commun. 11, 3509 (2020).

\bibitem{Jorgensen2020} 
P. B. Jørgensen and A. Bhowmik, DeepDFT: Neural message passing network for accurate charge density prediction, arXiv:2011.03346 (2020).

\bibitem{Jorgensen2022} 
P. B. Jørgensen and A. Bhowmik, Equivariant graph neural networks for fast electron density estimation of molecules, liquids, and solids, npj Comput. Mater. 8, 183 (2022).

\bibitem{Lee2022} 
A. J. Lee, J. A. Rackers, and W. P. Bricker, Predicting accurate ab initio DNA electron densities with equivariant neural networks, Biophys. J. 121, 3883-3895 (2022).

\bibitem{Kim2024} 
S. Kim and S. Ahn, Gaussian plane-wave neural operator for electron density estimation, arXiv:2402.04278 (2024).

\bibitem{Pathrudkar2024} 
S. Pathrudkar, P. Thiagarajan, S. Agarwal, A. S. Banerjee, and S. Ghosh, Electronic structure prediction of multi-million atom systems through uncertainty quantification enabled transfer learning, npj Comput. Mater. 10, 175 (2024).

\bibitem{Koker2024} 
T. Koker, K. Quigley, E. Taw, K. Tibbetts, and L. Li, Higher-order equivariant neural networks for charge density prediction in materials, npj Comput. Mater. 10, 161 (2024).

\bibitem{Lewis2021} 
A. M. Lewis, A. Grisafi, M. Ceriotti, and M. Rossi, Learning Electron Densities in the Condensed Phase, J. Chem. Theory Comput. 17, 7203-7214 (2021).

\bibitem{Fu2024} 
X. Fu, A. Rosen, K. Bystrom, R. Wang, A. Musaelian, B. Kozinsky, T. Smidt, and T. Jaakkola, A recipe for charge density prediction, Adv. Neural Inf. Process. Syst. 37, 9727-9752 (2024).

\bibitem{Lee2024}
R. G. Lee, Y. H. Kim Convolutional network learning of self-consistent electron density via grid-projected atomic fingerprints. npj Comput Mater 10, 248 (2024).

\bibitem{Cheng2023} 
C. Cheng and J. Peng, Equivariant neural operator learning with graphon convolution, [EB/OL]. (2023).

\bibitem{Kohn1996}
W. Kohn, Density Functional and Density Matrix Method Scaling Linearly with the Number of Atoms, Phys. Rev. Lett. 76, 3168 (1996).

\bibitem{Andersen1995}
O. K. Andersen, O. Jepsen, and G. Krier, Exact muffin-tin orbital theory, in \textit{Lectures on Methods of Electronic Structure Calculations}, edited by V. Kumar, O. K. Andersen, and A. Mookerjee (World Scientific, Singapore, 1995), pp. 63–124.

\bibitem{Andersen2000}
O. K. Andersen, T. Saha-Dasgupta, R. W. Tank, C. Arcangeli, O. Jepsen, and G. Krier, Developing the MTO formalism, in \textit{Electronic Structure and Physical Properies of Solids}, edited by H. Dreyssé (Springer Berlin Heidelberg, Berlin, Heidelberg, 2000), pp. 3–84.

\bibitem{Andersen2000PRB}
O. K. Andersen and T. Saha-Dasgupta, Muffin-tin orbitals of arbitrary order, Phys. Rev. B 62, R16219 (2000).


\bibitem{Vitos2007}
L. Vitos, \textit{Exact muffin-tin orbitals method, in Computational Quantum Mechanics for Materials Engineers: The EMTO Method and Applications} (Springer, London, 2007), pp. 13–35.

\bibitem{ZhangQ2019}
Q. Zhang, J. Yan, Y. Zhang, and Y. Ke, Exact muffin tin or- bital based first-principles method for electronic-structure and electron-transport simulation of device materials, Phys. Rev. B 100, 075134 (2019).

\bibitem{Chen2020}
Z. Chen, Q. Zhang, Y. Zhang, L. Wang, M. Sang, and Y. Ke, Exact muffin-tin orbital based fully relativistic simulation of device materials: Electronic charge and spin current, Phys. Rev. B 102, 035405 (2020).

\bibitem{ZhangY2021}
Y. Zhang, J. Zhai, Z. Chen, Q. Zhang, and Y. Ke, First-principles nonequilibrium dynamical cluster theory for quantum transport simulations of disordered nanoelectronic devices, Phys. Rev. B 104, 115412 (2021).

\bibitem{Chen2023}
Z. Chen, Q. Zhang, Z. Yuan, J. Liu, K. Xia, and Y. Ke, Fully relativistic first-principles quantum transport simulation of non- collinear spin transfer and spin hall current, Phys. Rev. B 107, 195431 (2023).

\bibitem{Turek1997}
I. Turek, V. Drchal, J. Kudrnovsky, M. Sob, and P. Weinberger, \textit{Electronic Structure of Disordered Alloys, Surfaces and Interfaces} (Springer, Boston, MA, 1997).

\bibitem{Pourovskii2005}
L. V. Pourovskii, A. V. Ruban, L. Vitos, H. Ebert, B. Johansson, and I. A. Abrikosov, Fully relativistic spin-polarized exact muffin-tin-orbital method, Phys. Rev. B 71, 094415 (2005).

\bibitem{ZhangA2024}
A. Zhang, Q. Zhang, Y. Zhang, Z. Chen, Y. Wu, and Y. Ke, Full-potential screened spherical wave based muffin-tin orbital method for density functional theory, Phys. Rev. B 110, 155126 (2024).

\bibitem{ZhangA2025}
A. Zhang, Q. Zhang, Z. Chen, Y. Wu, and Y. Ke, Implementation of the full-potential screened spherical wave-based muffin-tin orbital method for all-electron density functional theory, Phys. Rev. B 111, 235137 (2025).

\bibitem{TangM2024}
M. Tang, C. Liu, A. Zhang, Q. Zhang, J. Zhai, S. Yuan, and Y. Ke, Random Green’s function method for large-scale electronic structure calculation, Chin. Phys. Lett. 41, 053102 (2024), Express.

\bibitem{ZhangQ2024}
Q. Zhang, M. Tang, L. Wang, K. Xia, and Y. Ke, Random nonequilibrium Green’s function method for large- scale quantum transport simulation, Phys. Rev. B 110, 155430 (2024).

\bibitem{Datta2005}
S. Datta, \textit{Quantum Transport: Atom to Transistor} (Cambridge University Press, Cambridge, 2005).

\bibitem{Pashov2020}
D. Pashov, S. Acharya, W. R. Lambrecht, J. Jackson, K. D. Belashchenko, A. Chantis, F. Jamet, and M. van Schilfgaarde, Full charge-density calculation of the surface energy of metals, Comput. Phys. Commun. 249, 107065 (2020).

\bibitem{Kresse1999}
G. Kresse and D. Joubert, From ultrasoft pseudopotentials to the projector augmented-wave method, Phys. Rev. B 59, 1758 (1999).



\bibitem{Ferreira2008}
L. G. Ferreira, M. Marques, and L. K. Teles, Approximation to density functional theory for the calculation of band gaps of semiconductors, Phys. Rev. B 78, 125116 (2008).

\bibitem{Ferreira2011}
L. G. Ferreira, M. Marques, and L. K. Teles, Slater half-occupation technique revisited: the LDA-1/2 and GGA-1/2 approaches for atomic ionization energies and band gaps in semiconductors, AIP ADVANCES 1, 032119 (2011).

\bibitem{Kane1957}
E. O. Kane, Band structure of indium antimonide, J. Phys. Chem. Solids 1, 249–261 (1957).

\bibitem{Hansen1982}
G. L. Hansen, J. Schmit, and T. Casselman, Energy gap versus alloy composition and temperature in Hg$_{1-x}$Cd$_x$Te. J. Appl. Phys. 53, 7099-7101 (1982).




\end{thebibliography}
\end{document}